\documentstyle[12pt]{article}
\newcommand{\eV}{~\mbox{eV}}
\newcommand{\keV}{~\mbox{keV}}
\newcommand{\MeV}{~\mbox{MeV}}
\newcommand{\GeV}{~\mbox{GeV}}
\newcommand{\TeV}{~\mbox{TeV}}

\newcommand{\gsim}{ \mathop{}_{\textstyle \sim}^{\textstyle >} }
\newcommand{\lsim}{ \mathop{}_{\textstyle \sim}^{\textstyle <} }
\newcommand{\vev}[1]{ \left\langle {#1} \right\rangle }
\newcommand{\sneu}{\widetilde{N_1}}
\def\Frac#1#2{{\displaystyle\frac{#1}{#2}}}
\begin{document}
\baselineskip 0.6cm
\setcounter{footnote}{1}

\begin{titlepage}

\begin{flushright}
UT-957\\
LBNL-48679\\
UCB-PTH-01/30\\
hep-ph/0109030\\
\end{flushright}

\vskip 1cm
\begin{center}
{\large \bf Leptogenesis from $\widetilde{N}$-dominated early universe}

\vskip 1.2cm 

K.~Hamaguchi$^1$, Hitoshi Murayama$^{2,3}$ and T. Yanagida$^{1,4}$
\vskip 0.4cm

{\it $^1$ Department of Physics\\
University of Tokyo, Tokyo 113-0033, Japan}\\
{\it $^2$ Theoretical Physics Group\\
Ernest Orlando Lawrence Berkeley National Laboratory, MS 50A-5101\\
University of California, Berkeley, California 94720}\\
{\it $^3$ Department of Physics\\
University of California, Berkeley, California 94720}\\
{\it $^4$ Research Center for the Early Universe\\
University of Tokyo,
Tokyo 113-0033, Japan}
\vskip 2cm
\abstract{We investigate in detail the leptogenesis by the decay of
  coherent right-handed sneutrino $\widetilde{N}$ having dominated the
  energy density of the early universe, which was originally proposed by
  HM and TY. Once the $\widetilde{N}$ dominant universe is realized, the
  amount of the generated lepton asymmetry (and hence baryon asymmetry)
  is determined only by the properties of the right-handed neutrino,
  regardless of the history before it dominates the universe. Moreover,
  thanks to the entropy production by the decay of the right-handed
  sneutrino, thermally produced relics are sufficiently diluted. In
  particular, the cosmological gravitino problem can be avoided even
  when the reheating temperature of the inflation is higher than
  $10^{10}\GeV$, in a wide range of the gravitino mass $m_{3/2}\simeq
  10\MeV$--$100\TeV$.  If the gravitino mass is in the range
  $m_{3/2}\simeq 10\MeV$--$1\GeV$ as in the some gauge-mediated
  supersymmetry breaking models, the dark matter in our universe can be
  dominantly composed of the gravitino.  Quantum fluctuation of the
  $\widetilde{N}$ during inflation causes an isocurvature fluctuation
  which may be detectable in the future.}

\end{center}
\end{titlepage}

\setcounter{footnote}{0}

\section{Introduction}

Neutrino oscillations, especially the atmospheric neutrino oscillation
observed in the SuperKamiokande experiments~\cite{SK-Atm}, is one of the
greatest discoveries in the field of particle physics after the success
of the standard model.  The data suggest small but finite masses of the
neutrinos.  Such small neutrino masses can be naturally obtained via the
seesaw mechanism~\cite{seesaw} implying the existence of the lepton
number violation.  There has been, therefore, growing interest in
leptogenesis~\cite{LGorg} as a production mechanism of the baryon
asymmetry in the present universe. In fact, the ``sphaleron''
process~\cite{sphaleron} converts the lepton asymmetry into the baryon
asymmetry, and nonzero lepton asymmetry can be produced by the decay of
the heavy right-handed neutrino~\cite{LGorg}.

On the other hand, the supersymmetry (SUSY) has been regarded as an
attractive candidate for physics beyond the standard model, since it
protects the huge hierarchy between the electroweak and unification
scales against the radiative corrections as well as leads to a beautiful
unification of the gauge coupling constants. In Ref.~\cite{MY}, HM and
TY proposed new possibilities for leptogenesis in the framework of the
SUSY. Under the assumption of the SUSY, there appears a very simple and
attractive mechanism to produce the lepton asymmetry,\footnote{Another
interesting possibility for leptogenesis with SUSY proposed in
Ref.~\cite{MY} is the leptogenesis via the flat direction including the
charged lepton doublet $L$~\cite{MY,MM,AFHY}, which is based on the
Affleck-Dine mechanism~\cite{AD}.} that is, the condensation of the
scalar component of the right-handed neutrino and its decay into the
leptons and anti-leptons.

In this paper, we investigate in detail the leptogenesis by the decay of
a coherent right-handed sneutrino. In particular, we discuss the case in
which the coherent oscillation of the right-handed sneutrino dominates
the energy density of the early universe. It is extremely interesting
that the amount of produced baryon asymmetry is determined mainly by the
decay rate of the right-handed neutrino, whatever happened before the
coherent oscillation dominates the universe. Furthermore, as a big
bonus, thermally produced gravitinos are diluted by the entropy
production due to the decay of the coherent right-handed sneutrino, so
that the cosmological gravitino
problem~\cite{Gprob,Gprob-heavy,Gprob-1keV,Gprob-GMSB} can be avoided
even when the reheating temperature $T_R$ of the inflation is higher
than $10^{10}\GeV$, in a wide range of the gravitino mass $m_{3/2}\sim
10\MeV$--$100\TeV$.

In particular, this dilution of the thermally produced gravitinos has
great advantages in the gauge-mediated SUSY breaking (GMSB)
models~\cite{GMSB}. The GMSB mechanism has been regarded as a very
attractive candidate for the SUSY breaking, since it suppresses quite
naturally the flavor changing processes, which are inherent problems in
the SUSY standard model. In general, GMSB models predict that the
gravitino is the lightest SUSY particle~\footnote{This is not the case
if the SUSY breaking is mediated by a bulk gauge field in higher
dimension spacetime~\cite{bulkU1}.} and stable.\footnote{We assume here
that the $R$-parity is exact.}  Usually, the relic abundance of the
gravitino is proportional to the reheating temperature, and there are
severe upper bounds on the reheating temperature $T_R$ depending on the
gravitino mass $m_{3/2}$, in order to avoid that the energy density of
the gravitino overclose the present
universe~\cite{Gprob-GMSB}.\footnote{For a very light gravitino
$m_{3/2}\lsim 1\keV$, there is no gravitino problem~\cite{Gprob-1keV}.}
In our scenario, however, this overclosure bound is completely removed
because of the aforementioned right-handed sneutrino decay, and a
reheating temperature even higher than $10^{10}\GeV$ is possible for
$m_{3/2}\gsim 10\MeV$. Furthermore, as we will see, the present energy
density of the gravitino is determined independently of the reheating
temperature, and the gravitino mass can be {\it predicted} as
$m_{3/2}\simeq 10\MeV$--$1\GeV$ from the baryon asymmetry in the present
universe, if the dominant component of the dark matter is the gravitino.

\section{Leptogenesis by coherent right-handed sneutrino}
\subsection{The MSSM With Right-handed Neutrinos}

Let us start by introducing three generations of heavy right-handed
neutrinos $N_i$ with masses $M_i$ to the minimal supersymmetric standard
model (MSSM), which have a superpotential;
\begin{eqnarray}
\label{eq-super}
W = \frac{1}{2} M_i N_i N_i + h_{i\alpha} N_i L_{\alpha} H_u
\,,
\end{eqnarray}
where $L_{\alpha}$ ($\alpha = e, \mu, \tau$) and $H_u$ denote the
supermultiplets of the lepton doublets and the Higgs doublet which
couples to up-type quarks, respectively. The small neutrino mass is
obtained by integrating out the heavy right-handed neutrinos, which is
given by~\cite{seesaw}
\begin{eqnarray}
 (m_{\nu})_{\alpha\beta}
  = 
  - \sum_i h_{i\alpha} h_{i\beta}
  \frac{\vev{H_u}^2}{M_i}
  \,.
  \label{eq_seesaw}
\end{eqnarray}

During inflation, the scalar component of the right-handed neutrino
$\widetilde{N}$ can acquire a large amplitude~\cite{MY,DRT,GMO} if the
Hubble expansion rate of the inflation $H_{\rm inf}$ is larger than the
mass of the $\widetilde{N}$. Let us assume that there exists (at least)
one right-handed neutrino with a mass lighter than $H_{\rm inf}$, and
that it develops a large expectation value during the
inflation. Hereafter, we focus on the lightest right-handed sneutrino
$\sneu$ for simplicity. (Possible contributions from the heavier
right-handed sneutrinos $\widetilde{N_2}$ and $\widetilde{N_3}$ will be
discussed at the end of this section.)  It is assumed here that the
potential for the right-handed neutrino is given simply by the mass term
\begin{equation}
  V = M_1^2 |\widetilde{N}_1|^2
\end{equation}
and $L$ and $H_u$ vanish.\footnote{The parameters we prefer (as we will
  see later) give a large effective mass to $L$ and $H_u$ because
  $\widetilde{N}_1 \sim M_{pl}$.  Therefore, vanishing $L$ and $H_u$ is
  natural.}

After the end of the inflation, the Hubble parameter $H$ decreases with
cosmic time $t$ as $H\propto t^{-1}$, and $\sneu$ begins to oscillate
around the origin when $H$ becomes smaller than the mass of the
right-handed sneutrino $M_1$.  Then, the coherent oscillation eventually
decays when $H = \Gamma_{N_1}$ ($t\sim \Gamma_{N_1}^{-1}$), where
$\Gamma_{N_1} = (1/4\pi)\sum_{\alpha} |h_{1\alpha}|^2 M_1$ is the decay
rate of the $\sneu$. Because $\sneu$ decays into leptons (and Higgs) as
well as their anti-particles, its decay can produce lepton-number
asymmetry if $CP$ is not conserved~\cite{LGorg}. The generated lepton
number density is given by
\begin{eqnarray}
\label{eq-nL}
n_L = \epsilon_1 M_1 |\sneu_d|^2
\,,
\end{eqnarray}
where $|\sneu_d|$ is the amplitude of the oscillation when it decays,
and $\epsilon_1$ denotes the lepton-asymmetry parameter in the decay of
$\sneu$. Assuming a mass hierarchy $M_1 \ll M_2, M_3$ in the
right-handed neutrino sector, the explicit form of $\epsilon_1$ is given
by~\cite{ep-calc}
\begin{eqnarray}
\epsilon_1 
&\equiv& 
\frac{
\Gamma (\sneu \to L + H_u) 
-
\Gamma (\sneu \to \overline{L} + \overline{H_u})
}{
\Gamma (\sneu \to L + H_u) 
+
\Gamma (\sneu \to \overline{L} + \overline{H_u})
}
\nonumber
\\
&\simeq&
- \frac{3}{16\pi}
\frac{1}{\left(h h^{\dagger}\right)_{11}}
\sum_{i = 2,3}
{\rm Im} 
\left[
\left(h h^{\dagger}\right)_{1i}^2
\right]
\frac{M_1}{M_i}
\,.
\end{eqnarray}
Here, $L$ and $H_u$ ($\overline{L}$ and $\overline{H_u}$) symbolically
denote fermionic or scalar components of corresponding supermultiplets
(and their anti-particles).  By using the seesaw formula in
Eq.(\ref{eq_seesaw}), this $\epsilon_1$ parameter can be rewritten in
terms of the heaviest neutrino mass $m_{\nu_3}$ and an effective $CP$
violating phase $\delta_{\rm eff}$~\cite{BY};
\begin{eqnarray}
\label{eq-eps1}
\epsilon_1
&=&
\frac{3}{16\pi}
\frac{M_1}{\vev{H_u}^2}
\frac{
{\rm Im}\left[h (m_{\nu}^*) h^T\right]_{11}
}
{
\left(h h^{\dagger}\right)_{11}
}
\nonumber \\
&\equiv&
\frac{3}{16\pi}
\frac{M_1}{\vev{H_u}^2}
m_{\nu_3}
\delta_{\rm eff}
\nonumber \\
&\simeq&
1 \times 10^{-10}
\left(
\frac{M_1}{10^6\GeV}
\right)
\left(
\frac{m_{\nu_3}}{0.05\eV}
\right)
\delta_{\rm eff}
\,.
\end{eqnarray}
Here, we have used $\vev{H_u} = 174\GeV \times \sin\beta$, where
$\tan\beta\equiv \vev{H_u}/\vev{H_d}$. ($H_d$ is the Higgs field which
couples to down-type quarks.) Here and hereafter, we take
$\sin\beta\simeq 1$ for simplicity. As for the heaviest neutrino mass,
we take $m_{\nu_3}\simeq 0.05\eV$ as a typical value, suggested from the
atmospheric neutrino oscillation observed in the SuperKamiokande
experiments~\cite{SK-Atm}.

\subsection{Cosmic Lepton Asymmetry}

The fate of the generated lepton asymmetry depends on whether or not the
coherent oscillation of $\sneu$ dominates the energy density of the
universe before it decays~\cite{MY}.  In this paper, we mainly discuss
the leptogenesis scenario from the universe dominated by $\sneu$. (We
will give a brief comment on the case where $\sneu$ does not dominate
the universe in Appendix.) As we shall show soon, once the $\sneu$
dominant universe is realized, the present baryon asymmetry is
determined only by the properties of the right-handed neutrino, whatever
happened before the $\sneu$ dominates the universe.  We first derive the
amount of the generated lepton asymmetry just assuming that the $\sneu$
dominates the universe, and after that we will discuss the necessary
conditions of the present scenario.

Once $\sneu$ dominates the universe before it decays, the universe is
reheated again at $H = \Gamma_{N_1}$ by the decay of $\sneu$.  The
energy density of the resulting radiation, with a temperature $T_{N_1}$,
is given by the following relation;
\begin{eqnarray}
\label{eq-TN}
\frac{\pi^2}{30} g_* T_{N_1}^4
&=&
M_1^2 |\sneu_d|^2
\nonumber \\
&=&
3 M_{pl}^2 \Gamma_{N_1}^2
\,,
\end{eqnarray}
while the entropy density is given by
\begin{eqnarray}
\label{eq-s}
s = \frac{2\pi^2}{45} g_* T_{N_1}^3
\,.
\end{eqnarray}
Here, $M_{pl} = 2.4\times 10^{18}\GeV$ is the reduced Planck scale and
$g_*$ is the number of effective degrees of freedom, which is $g_*
\simeq 200$ for temperatures $T\gg 1\TeV$ in the SUSY standard model.
{}From the above equations, the ratio of the lepton number density to
the entropy density is given by the following simple form;
\begin{eqnarray}
\label{eq-nL-s}
\frac{n_L}{s} 
&=& 
\frac{3}{4}\epsilon_1 \frac{T_{N_1}}{M_1}
\nonumber \\
&\simeq& 0.7 \times 10^{-10}
\left(
\frac{T_{N_1}}{10^6\GeV}
\right)
\left(
\frac{m_{\nu_3}}{0.05\eV}
\right)
\delta_{\rm eff}
\,.
\end{eqnarray}
Here, we have required that the decay of the $\sneu$ occurs in an
out-of-equilibrium way, namely, $T_{N_1} < M_1$, so that the produced
lepton-number asymmetry not be washed out by lepton-number violating
interactions mediated by $N_1$.

Because the lepton asymmetry is produced before the electroweak phase
transition at $T\sim 100\GeV$, it is partially converted~\cite{LGorg}
into the baryon asymmetry through the ``sphaleron''
effects~\cite{sphaleron};
\begin{eqnarray}
\frac{n_B}{s} = a \frac{n_L}{s}
\,,
\end{eqnarray}
where $a = -8/23$ in the SUSY standard model~\cite{L-to-B}. This ratio
takes a constant value as long as an extra entropy production does not
take place at a later epoch. Therefore, as mentioned in the
introduction, the baryon asymmetry in the present universe is indeed
determined only by the decay temperature of the right-handed sneutrino
$T_{N_1}$ (and the effective $CP$ violating phase $\delta_{\rm eff}$),
given in Eq.(\ref{eq-nL-s}). Thus it is independent of unknown
parameters of the inflation such as the reheating temperature
$T_R$. Assuming the effective $CP$ violating phase $\delta_{\rm eff}
\,(\le 1)$ to be not too small, the observed baryon asymmetry $n_B/s
\simeq (0.4$--$1)\times 10^{-10}$~\cite{KT} is obtained by taking
\begin{eqnarray}
\label{eq-TN-required}
T_{N_1} \simeq 10^6 - 10^7\GeV
\,.
\end{eqnarray}

Now let us recall the conditions we have required so far. We have
required the following two conditions; (i) $\sneu$ dominates the
universe before it decays, and (ii) $\sneu$ decays in an
out-of-equilibrium way.  By taking the $T_{N_1}$ in
Eq.(\ref{eq-TN-required}), the condition of the out-of-equilibrium decay
is given by
\begin{eqnarray}
\label{eq-mass}
M_1 > T_{N_1} \simeq 10^6 - 10^7\GeV
\,.
\end{eqnarray}
Notice that the temperature $T_{N_1}$ is determined by the decay rate of
the $\sneu$ [see Eq.(\ref{eq-TN})], and hence is related to the mass and
couplings of $\sneu$. The relation is given by
\begin{eqnarray}
\label{eq-coupling}
\sqrt{\sum_{\alpha} |h_{1\alpha}|^2} 
\simeq 5\times 10^{-6}
\left( \frac{T_{N_1}}{10^6\GeV} \right)^{1/2} 
\left( \frac{T_{N_1}}{M_1}\right)^{1/2} 
\,.
\end{eqnarray}
Thus, we need Yukawa couplings $h_{1\alpha}$ which are as small as the
electron Yukawa coupling.

\subsection{Conditions For $\tilde{N}$-dominance}

In order to discuss whether or not $\sneu$ dominates the universe, it is
necessary to consider the history of the universe before it
decays. Here, we assume that the potential of the $\sneu$ is ``flat'' up
to the Planck scale, namely, the potential is just given by the mass
term $M_1^2 |\sneu|^2$ up to the Planck scale. (This may not be the case
when the masses of the right-handed neutrinos are induced by a breaking
of an additional gauge symmetry. We will discuss such a case in the next
section.)

Assuming the flatness of the $\sneu$'s potential up to the Planck scale
({\it i.e.}\/, only the mass term), the initial amplitude of the
oscillation is naturally given by $|\sneu_i| \simeq M_{pl}$, since above
the Planck scale the scalar potential is expected to be exponentially
lifted by the supergravity effects.\footnote{Even though it is possible
that $\sneu$ has a larger initial amplitude $|\sneu_i| > M_{pl}$ (see,
e.g., Ref~\cite{MSYY}), it depends on the scalar potential beyond the
Planck scale, so that we do not discuss this possibility in this paper.} 
Then, the energy density of $\sneu$ when it starts the coherent
oscillation is given by $\rho_{N_1} \simeq M_1^2 M_{pl}^2$.

The rest of the total energy density of the universe at $H = M_1$ is
dominated by (i) the oscillating inflaton $\psi$ or (ii) the radiation,
depending on the decay rate of the inflaton $\Gamma_{\psi}$. If
$\Gamma_{\psi} < M_1$, the reheating process of the inflation has not
completed yet at $H = M_1$, and the inflaton $\psi$ is still oscillating
around its minimum, whose energy density is given by $\rho_{\psi} \simeq
2 M_1^2 M_{pl}^2$.  The ratio of the energy density of $\sneu$ to that
of the inflaton, $\rho_{N_1} / \rho_{\psi} \simeq 1 / 2$, takes a
constant value until either of these oscillations decays. Because the
energy density of the radiation $\rho_{\rm rad}$ resulting from the
inflaton decay is diluted faster than $\rho_{N_1}$, the oscillating
$\sneu$ dominates the universe if its decay rate $\Gamma_{N_1}$ is slow
enough compared with that of the inflaton $\Gamma_{\psi}$; $\Gamma_{N_1}
\ll \Gamma_{\psi}$.

On the other hand, if $\Gamma_{\psi} > M_1$, the inflaton decay has
already completed before $H = M_1$, and the energy density of the
radiation at $H = M_1$ is given by $\rho_{\rm rad} \simeq 2 M_1^2
M_{pl}^2$. In this case, the oscillating $\sneu$ dominates the universe
soon after it starts the oscillation and hence before its
decay.\footnote{This is the case as long as $\Gamma_{N_1} \ll M_1.$}
Therefore, the condition for $\sneu$ to dominate the universe is just
given by $\Gamma_{N_1} \ll \Gamma_{\psi}$. In terms of the reheating
temperature $T_R$, it is
\begin{eqnarray}
\label{eq-dom-cond}
T_R \gg  T_{N_1}
\simeq 10^6 - 10^7\GeV
\,,
\end{eqnarray}
which is easily satisfied in various SUSY inflation
models~\cite{S-inflations}.  Thus, the present leptogenesis scenario
from $\sneu$ dominated early universe is almost automatic as long as the
right-handed neutrino has suitable mass and couplings given in
Eqs.(\ref{eq-mass}) and (\ref{eq-coupling}).

\subsection{Gravitino Problem Ameliorated}

Now let us turn to consider the cosmological gravitino
problem~\cite{Gprob,Gprob-heavy,Gprob-1keV,Gprob-GMSB}. There are two
cases; unstable and stable gravitino. When the gravitino is not the
lightest SUSY particle, it has a very long lifetime, and its decay
during or after the Big-Bang Nucleosynthesis (BBN) epoch ($t\sim
1$--$100$ sec) might spoil the success of the BBN.  Since the abundance
of the thermally produced gravitinos at reheating epoch is proportional
to the reheating temperature $T_R$, usually there are upper bounds on
the $T_R$ depending on the gravitino mass. The bound is given by $T_R
\lsim 10^7$--$10^9\GeV$ for $m_{3/2} \simeq
100\GeV$--$1\TeV$~\cite{Gprob}, and $B_h T_R \lsim 10^7$--$10^9\GeV$ for
$m_{3/2}\simeq$ (a few -- $100)\TeV$~\cite{Gprob-heavy}, where $B_h$
denotes branching ratio of the gravitino decay into hadrons. However, in
the present scenario, the gravitino abundance is diluted by the entropy
production due to the right-handed sneutrino decay. The dilution factor
is given by
\begin{eqnarray}
\Delta 
= 
\left\{
\begin{array}{lc}
\Frac{T_R}{2\,T_{N_1}} 
&
({\rm for}\quad T_R < T_{R_C})
\\
\Frac{T_{R_C}}{2\,T_{N_1}} 
&
({\rm for}\quad T_R > T_{R_C})
\end{array}
\right.
\,,
\end{eqnarray}
where 
\begin{eqnarray}
T_{R_C} \equiv 7\times 10^{11}
\left(\frac{M_1}{10^6\GeV}\right)^{1/2} \GeV
\,. 
\end{eqnarray}
Here, $T_R < T_{R_C}$ ( $T_R > T_{R_C}$ ) corresponds to $\Gamma_{\psi}
< M_1$ ( $\Gamma_{\psi} > M_1$ ). Thanks to this entropy production by
the $\sneu$'s decay, the constraint from the gravitino problem applies
not to the reheating temperature $T_R$, but to an effective temperature
given by
\begin{eqnarray}
\label{eq-Tgrav}
T_{R\,{\rm eff}} \equiv \frac{1}{\Delta} T_R 
&=& 
\left\{
\begin{array}{l}
2\, T_{N_1}
\\
2\, T_{N_1}
\left(
\Frac{T_R}{T_{R_C}}
\right)
\end{array}
\right.
\nonumber
\\
&\simeq& 
\left\{
\begin{array}{lc}
2 \times 10^6 - 2\times 10^7\GeV
&
({\rm for}\quad T_R < T_{R_C})
\\
2\times 10^6 - 2\times 10^7\GeV 
\times
\left(
\Frac{T_R}{T_{R_C}}
\right)
&
({\rm for}\quad T_R > T_{R_C})
\end{array}
\right.
\,,
\end{eqnarray}
which is much below the original reheating temperature $T_R$. Therefore,
the cosmological gravitino problem can be avoided in a wide range of the
gravitino mass $m_{3/2}\simeq 100\GeV$--$100\TeV$, even if the reheating
temperature $T_R$ of the inflation is higher than $10^{10}\GeV$. The
fact that such high reheating temperature is allowed makes it very easy
to construct realistic SUSY inflation models.

On the other hand, if the gravitino is the lightest SUSY particle, as in
the GMSB scenario, it is completely stable. If there is no extra entropy
production after the inflation, the relic abundance of the gravitinos
which are produced thermally after the inflation is given
by~\cite{Gprob-GMSB}
\begin{eqnarray}
 \left.
  \Omega_{3/2}\,h^2
  \right|_{{\rm without}\,\,\sneu\,\,{\rm decay}}
  \simeq
  0.8\times
  \left(
   \frac{M_3}
   {1\TeV}
   \right)^2
   \left(
   \frac{m_{3/2}}
   {10\MeV}
   \right)^{-1}
    \left(
     \frac{T_R}
     {10^6\GeV}
     \right)
     \,.
     \label{omega32-org}
\end{eqnarray}
Here, $M_3$ is the gluino mass, $h$ is the present Hubble parameter in
units of $100$ km sec$^{-1}$\,Mpc$^{-1}$ and $\Omega_{3/2} =
\rho_{3/2}/\rho_c$ .  ($\rho_{3/2}$ and $\rho_c$ are the present energy
density of the gravitino and the critical energy density of the present
universe, respectively.) It is found from Eq.(\ref{omega32-org}) that
the overclosure limit $\Omega_{3/2} < 1$ puts a severe upper bound on
the reheating temperature $T_R$, depending on the gravitino mass
$m_{3/2}$.  However, in our scenario, the ``reheating'' by the coherent
$\sneu$ takes place and the relic abundance of the gravitino is obtained
by dividing the original abundance in Eq.(\ref{omega32-org}) by the
dilution factor $\Delta$:
\begin{eqnarray}
 \left.
  \Omega_{3/2}\,h^2
  \right|_{{\rm with}\,\,\sneu\,\,{\rm decay}}
  &\simeq&
  \frac{1}{\Delta}
  \left.
   \Omega_{3/2}\,h^2
   \right|_{{\rm without}\,\,\sneu\,\,{\rm decay}}
   \nonumber \\
 &\simeq&
  0.8\times
  \left(
   \frac{M_3}
   {1\TeV}
   \right)^2
   \left(
   \frac{m_{3/2}}
   {10\MeV}
   \right)^{-1}
    \left(
     \frac{T_{R\,{\rm eff}}}
     {10^6\GeV}
     \right)
     \,.
\end{eqnarray}
Therefore, again, the overclosure problem can be avoided almost
independently of the reheating temperature $T_R$, and a reheating
temperature even higher than $10^{10}\GeV$ is possible for $m_{3/2}
\gsim 10\MeV$. Moreover, it is found from this equation that the present
energy density of the gravitino is independent of the reheating
temperature, in a very wide range of $T_{N_1} < T_R < T_{R_C}$. Thus, we
can predict the gravitino mass by requiring that the gravitino is the
dominant component of the dark matter;
\begin{eqnarray}
 m_{3/2}
  &\simeq&
  50\MeV
  \times
  \left(
   \frac{M_3}
   {1\TeV}
   \right)^2
   \left(
    \frac{\Omega_{\rm matter}\,h^2}
    {0.15}
    \right)^{-1}
    \left(
     \frac{T_{R\,{\rm eff}}}
     {10^6\GeV}
     \right)
     \nonumber \\
 &\simeq&
  100\MeV - 1\GeV
  \times
  \left(
   \frac{M_3}
   {1\TeV}
   \right)^2
   \left(
    \frac{\Omega_{\rm matter}\,h^2}
    {0.15}
    \right)^{-1}
    \,,
\end{eqnarray}
for $T_{N_1} < T_R < T_{R_C}$.\footnote{One might wonder if the decay of
the next-to-lightest SUSY particle into gravitino during or after the
BBN would spoil the success of the BBN in the GMSB scenario. However,
this problem is avoided for $m_{3/2}\lsim 1\GeV$~\cite{MMY,LSPgrav}.} 
Here, we take the present matter density $\Omega_{\rm matter}\simeq 0.3$
and $h \simeq 0.7$~\cite{PDB}. Notice that this prediction comes from
the fact that the present energy density of the gravitino is determined
by the effective temperature $T_{R\,{\rm eff}} = 2\,T_{N_1}$ (for $T_R <
T_{R_C}$), while the decay temperature of the right-handed neutrino
$T_{N_1}$ is fixed by the baryon asymmetry in the present universe (see
Eq.(\ref{eq-TN-required})).

\subsection{Some Discussions}

Before closing this section, several comments are in order. The first
one is about the neutrino mass $m_{\nu}$. The contribution to the
neutrino mass matrix $(m_{\nu})_{\alpha\beta}$ from $N_1$ is given by
\begin{eqnarray}
 \label{eq-mnu-from1}
  \left|
   (m_{\nu})_{\alpha\beta}^{\rm from\,\,N_1}
   \right|
   &=&
   \left| h_{1\alpha} h_{1\beta}\right|
   \frac{\vev{H_u}^2}{M_1}
   \nonumber \\
 &\le&
  \sum_{\alpha}|h_{1\alpha}|^2
   \frac{\vev{H_u}^2}{M_1}
   \simeq
   7\times 10^{-4} \eV
   \left(\frac{T_{N_1}}{M_1}\right)^2
   \,.
\end{eqnarray}
Here, we have used the relation in Eq.(\ref{eq-coupling}). Therefore, it
is understood that the mass scale of the neutrinos suggested from the
atmospheric and solar neutrino oscillations, $m_{\nu}\sim
10^{-1}$--$10^{-3} \eV$, should be induced from the heavier right-handed
neutrinos, $N_2$ and $N_3$. The relative hierarchy between the mass and
couplings of $N_1$ and those of the $N_2$ and $N_3$ might be naturally
explained by a broken flavor symmetry.

For example, a broken discrete $Z_6$ symmetry~\cite{FHY} with a breaking
 parameter $\varepsilon \simeq 1/17$ and charges
 $Q(L_{\tau},L_{\mu},L_e) = (a,a,a+1)$ and $Q(N_3,N_2,N_1) = (b,c,3+d)$
 gives rise to the following superpotential;
\begin{eqnarray}
 W = \frac{1}{2} \sum_{(i,j)\ne (1,1)} g_{ij} 
 M_0 \,\varepsilon^{Q_i + Q_j} N_i N_j
 + \frac{1}{2}g_{11} M_0 \,\varepsilon^{2d} N_1 N_1
 + \tilde{h}_{i\alpha} \,\varepsilon^{Q_i + Q_{\alpha}} 
 N_i L_{\alpha} H_u
 \,,
\nonumber
\end{eqnarray}
where $g_{ij}$ and $\tilde{h}_{i\alpha}$ are ${\cal O}(1)$
couplings. The above charge assignments for lepton doublets naturally
lead to the realistic neutrino mass matrix including the maximal mixing
for the atmospheric neutrino oscillation~\cite{YanaRamo}. The overall
mass scale of the right-handed neutrino $M_0$ is determined by
$m_{\nu_3}\sim \varepsilon^{2a}\vev{H_u}^2/M_0$. By taking $a+d=2$, this
model gives $M_1\sim \varepsilon^{2d}M_0\sim 7\times 10^9\GeV$,
$\sqrt{\sum_{\alpha} |h_{1\alpha}|^2}\sim \varepsilon^5\sim 7\times
10^{-7}$, and hence $T_{N_1}\sim 1\times 10^7\GeV$.

So far, we have considered the leptogenesis from the lightest
right-handed sneutrino, $\sneu$.  The heavier right-handed sneutrino
$\widetilde{N_2}_{(3)}$ can also develop a large amplitude during the
inflation (if $M_{2(3)} < H_{\rm inf}$) and it may produce lepton
asymmetry in a similar way to the $\sneu$. However, the decay
temperatures of the $\widetilde{N_2}$ and $\widetilde{N_3}$ can not
satisfy the out-of-equilibrium condition $T_{2(3)} < M_1$, since $N_2$
and $N_3$ must explain the mass scales of the neutrino
oscillations. [See Eq.(\ref{eq-mnu-from1}).] Therefore, even if the
$\widetilde{N_2}_{(3)}$'s decay would produce additional lepton
asymmetry, it would be washed out and hence it can not contribute to the
resultant total lepton asymmetry.

Finally, we comment on the effects of the thermal
plasma~\cite{DRT,ACE,SomeIssues}, which might cause an early oscillation
of the right-handed sneutrino $\sneu$ before $H = M_1$. (Notice that
there is a dilute plasma with a temperature $T\simeq (T_R^2 M_{pl}
H)^{1/4}$ even before the reheating process of the inflation
completes~\cite{KT}.) There are basically two possible thermal
effects. First, when the temperature $T$ is higher than the effective
mass for $L$ and $H_u$, $T > m_{\rm eff} = \sqrt{\sum_{\alpha}
|h_{1\alpha}|^2} |\sneu|$, the $\sneu$ receives an additional thermal
mass $\delta M_{\rm th}^2 = (1/4)\sum_{\alpha} |h_{1\alpha}|^2 T^2$ from
the Yukawa coupling to $L$ and $H_u$~\cite{ACE}. Thus, the $\sneu$ field
would start an early oscillation if the additional thermal mass becomes
larger than the Hubble expansion rate before $H = M_1$. However, even if
$\sneu$ receives the thermal mass, the ratio of the thermal mass to the
Hubble expansion rate is given by
\begin{eqnarray}
\frac{\delta M_{\rm th}^2}{H^2}
\simeq 
\left\{
\begin{array}{lc}
0.07\times
\left(
\Frac{10 T_{N_1}}{M_1}
\right)^2
\left(
\Frac{M_1}{H}
\right)^{3/2}
\left(
\Frac{T_R}{T_{R_C}}
\right) 
&
{\rm for}\quad T_R < T_{R_C}
\,,
\\
0.03\times
\left(
\Frac{10 T_{N_1}}{M_1}
\right)^2
\left(
\Frac{M_1}{H}
\right)
&
{\rm for}\quad T_R > T_{R_C}
\,,
\end{array}
\right.
\end{eqnarray}
where we have used the relation given in
Eq.(\ref{eq-coupling}). Therefore, we can safely neglect the above
thermal effect, as long as $M_1$ is a bit larger than $T_{N_1}$. Next,
there is another thermal effect which has been pointed out in
Ref.~\cite{SomeIssues}. If the temperature is lower than the effective
mass for $L$ and $H_u$, $T < m_{\rm eff} = \sqrt{\sum_{\alpha}
|h_{1\alpha}|^2} |\sneu|$, the evolution of the running gauge and/or
Yukawa coupling constants $f(\mu)$ which couple to them are modified
below the scale $\mu = m_{\rm eff}$. Thus, these running coupling
constants depend on $|\sneu|$, and there appears an additional thermal
potential for $\sneu$;
\begin{eqnarray}
 \delta V(\sneu) = a T^4
  \log
  \left(
   \frac{|\sneu|^2}{T^2}
   \right)
   \,,
\end{eqnarray}
where $a$ is a constant of order ${\cal O}(f^4)$. However, again, it
turns out that the effective thermal mass for $\sneu$ is less than the
Hubble expansion rate;
\begin{eqnarray}
\label{eq-T4effect}
\frac{\delta M_{\rm th}^{'\, 2}}{H^2}
&=&
\frac{a T^4}{H^2 |\sneu|^2}
\nonumber\\
&\simeq& 
\left\{
\begin{array}{lc}
0.2\times
a
\left(
\Frac{M_{pl}}{|\sneu|}
\right)^2
\left(
\Frac{M_1}{H}
\right)
\left(
\Frac{T_R}{T_{R_C}}
\right)^2 
&
{\rm for}\quad T_R < T_{R_C}
\,,
\\
0.05\times
a
\left(
\Frac{M_{pl}}{|\sneu|}
\right)^2
&
{\rm for}\quad T_R > T_{R_C}
\,,
\end{array}
\right.
\end{eqnarray}
and hence this thermal effect is also irrelevant to the present scenario.

\section{Initial amplitude}
\label{sec-ini-amp}
In the previous section, we have assumed that the initial amplitude of
the $\sneu$'s oscillation is $|\sneu_i| \simeq M_{pl}$. This can be
realized when the right-handed neutrino has only the mass term up to the
Planck scale.  In this section, we discuss another possibility, where
the masses of the right-handed neutrinos are dynamically induced by a
spontaneously broken gauge symmetry. The simplest candidate is
U$(1)_{B-L}$, where $B$ and $L$ are baryon and lepton number,
respectively. Let us denote the chiral superfields whose vacuum
expectation values break the U$(1)_{B-L}$ by $\Phi$ and $\bar{\Phi}$.
(We need two fields with opposite charges in order to cancel
U$(1)_{B-L}$ gauge anomalies.) Due to the $D$-term and the $F$-term
coming from the superpotential which gives the right-handed neutrino
masses, the scalar potential of the right-handed sneutrino $\sneu$ is
lifted above the U$(1)_{B-L}$ breaking scale
$\vev{\Phi}$~\cite{MY}. Therefore, the initial amplitude of the
$\sneu$'s oscillation at $H\simeq M_1$ is given by $|\sneu_i|\sim
\vev{\Phi}$.

The breaking scale of the U$(1)_{B-L}$ gauge symmetry is model
dependent. If it is broken at the Planck scale, $\vev{\Phi}\simeq
M_{pl}$, the discussion in the previous section does not change at
all.\footnote{In this case, we need small couplings in order to explain
the intermediate right-handed neutrino mass scale. For example, a
superpotential $W = (1/2)y_i \Phi N_i N_i$ with $\vev{\Phi}\simeq
M_{pl}$ and $y_3 \sim 10^{-4}$ gives the mass $M_3 \sim 10^{14}\GeV$ to
the heaviest right-handed neutrino.  Such a small Yukawa coupling could
well be a consequence of broken flavor symmetries.}  On the other hand,
if $\vev{\Phi}$ is below the Planck scale, the initial amplitude of the
$\sneu$'s oscillation is reduced, and some parts of the discussion in
the previous section are modified; those are, the condition of the
$\sneu$ dominant universe [Eq.(\ref{eq-dom-cond})] and the effective
temperature of the cosmological gravitino problem
[Eq.(\ref{eq-Tgrav})].\footnote{For the reduced initial amplitude, the
thermal effect from the $aT^4\log(|\sneu|^2)$ potential becomes larger
than the case of $|\sneu_i|\simeq M_{pl}$. However, it is still
irrelevant for $a\sim {\cal O}(f^4)\lsim 10^{-2}$, as long as
$|\sneu_i|\gsim 10^{17}\GeV$. [See Eq.(\ref{eq-T4effect}).]} (Notice
that the amount of the generated lepton asymmetry given in
Eq.(\ref{eq-nL-s}) does not depend on the initial amplitude $|\sneu_i|$
as long as the $\sneu$ dominant universe is realized.) Let us take
$|\sneu_i|\sim \vev{\Phi}\sim 10^{17}\GeV$ for example. Due to the
reduced initial amplitude of $\sneu$, which means a smaller initial
energy density, the condition for $\sneu$ to dominate the universe is
now given by
\begin{eqnarray}
T_R 
&\gg&
T_{N_1}
\left(
\frac{\sqrt{3} M_{pl}}{|\sneu_i|}
\right)^2
\nonumber
\\
&\simeq&
2\times 10^9 - 2\times 10^{10}\GeV
\times
\left(
\frac{|\sneu_i|}{10^{17}\GeV}
\right)^{-2}
\,.
\end{eqnarray}
This condition is still easily satisfied by considering an inflation
with relatively high scale. On the other hand, the effective temperature
for the gravitino problem now becomes
\begin{eqnarray}
T_{R\,{\rm eff}}
&=& 
\left\{
\begin{array}{l}
T_{N_1}
\left(
\Frac{\sqrt{3} M_{pl}}{|\sneu_i|}
\right)^2
\\
T_{N_1}
\left(
\Frac{\sqrt{3} M_{pl}}{|\sneu_i|}
\right)^2
\left(
\Frac{T_R}{T_{R_C}}
\right)
\end{array}
\right.
\nonumber
\\
&=&
\left\{
\begin{array}{lc}
2\times 10^9 - 2\times 10^{10}\GeV
\times
\left(
\Frac{|\sneu_i|}{10^{17}\GeV}
\right)^{-2}
&
({\rm for}\quad T_R < T_{R_C})
\\
2\times 10^9 - 2\times 10^{10}\GeV
\times
\left(
\Frac{|\sneu_i|}{10^{17}\GeV}
\right)^{-2}
\left(
\Frac{T_R}{T_{R_C}}
\right)
&
({\rm for}\quad T_R > T_{R_C})
\end{array}
\right.
\,,
\nonumber \\
\end{eqnarray}
Thus, in this case, when the gravitino is unstable, its mass should be
in a range of $m_{3/2}\gsim 1\TeV$ to avoid the cosmological gravitino
problem. This difficulty might be avoided when the gravitino is stable
with mass $m_{3/2}\sim 10$--$100\GeV$~\cite{LSPgrav}.

\section{Discussion and Conclusions}

We have investigated in this paper leptogenesis from the universe
dominated by the right-handed sneutrino. We have found that this
scenario is very successful in explaining the present baryon
asymmetry. It is interesting that the amount of the generated lepton
asymmetry is determined mainly by the decay temperature of the
right-handed neutrino, independently of the reheating temperature $T_R$
of the inflation. The desirable amount of the baryon asymmetry in the
present universe is obtained when the decay temperature of the
right-handed neutrino is $T_{N_1}\simeq 10^6$--$10^7\GeV$.

An attractive feature of this scenario is the entropy production by the
decay of the coherent right-handed sneutrino, which itself produces the
lepton asymmetry. The abundance of the thermally produced gravitinos is
diluted by this entropy production, and the cosmological gravitino
problem can be avoided in a wide range of the gravitino mass
$m_{3/2}\simeq 10\MeV$--$100\TeV$. Actually, we have shown that the
effective temperature $T_{R\,{\rm eff}}$, to which the constraint from
the gravitino problem is applied, can be as low as $T_{R\,{\rm
eff}}\simeq 2\times 10^6$--$2\times 10^7\GeV$, even with such high
reheating temperatures as $T_R\gg 10^{10}\GeV$. The fact that such a
high reheating temperature is allowed is very welcome from the viewpoint
of building SUSY inflation models.

In particular, if the gravitino is stable, as in the gauge-mediated SUSY
breaking models, the present energy density of the gravitino is
determined by the decay temperature of the right-handed neutrino
$T_{N_1}\simeq 10^6$--$10^7\GeV$ (if we assume the initial amplitude of
the coherent right-handed sneutrino is $|\sneu_i| \simeq M_{pl}$). Thus,
the gravitino mass can be predicted from the observed energy density of
the dark matter as $m_{3/2}\simeq 10\MeV$--$1\GeV$, for a wide range of
the reheating temperature $10^6\GeV < T_R < 7\times 10^{11}
(M_1/10^6\GeV)^{1/2}\GeV$, assuming that the dark matter in our universe
is dominantly composed of the gravitino.

Finally, we comment on the isocurvature density perturbation coming from
the fluctuation of the initial amplitude of the right-handed sneutrino,
$|\sneu_i|$.\footnote{The authors thank M.~Kawasaki for useful
discussion.} The baryonic isocurvature perturbation from $\sneu$ is
given by
\begin{eqnarray}
 \delta_{\rm iso}
  &=&
  \frac{\delta n_B^{\rm iso}}{n_B}
  \frac{\Omega_B}{\Omega_t}
  \nonumber \\
 &\simeq&
  \frac{H_{\rm inf}}{\pi |\sneu_i|}
  \frac{\Omega_B}{\Omega_t}
  \nonumber \\
 &\simeq&
  1\times 10^{-7}
  \left(
   \frac{M_{pl}}{|\sneu_i|}
   \right)
   \left(
    \frac{H_{\rm inf}}{10^{13}\,{\rm GeV}}
    \right)
    \left(
     \frac{\Omega_B}{0.1\times \Omega_t}
     \right)
     \,,
\end{eqnarray}
where $\Omega_B$ and $\Omega_t$ is the density parameters of baryons and
total matter, respectively. This isocurvature fluctuation might be
detected in future experiments.

\section*{Acknowledgements}
HM and TY wish to express their thanks to M.~Kawasaki for discussion in
the early stage of the work. KH thanks the LBNL theory group for
hospitality, where part of this work has been done, and thanks M.~Fujii
and M.~Kawasaki for helpful discussions. He is supported by the Japanese
Society for the Promotion of Science. HM was supported in part by the
U.S.~Department of Energy under Contract DE-AC03-76SF00098, and in part
by the National Science Foundation under grant PHY-95-14797. TY
acknowledges partial support from the Grant-in-Aid for Scientific
Research from the Ministry of Education, Sports, and Culture of Japan,
on Priority Area \# 707: ``Supersymmetry and Unified Theory of
Elementary Particles''.

\appendix
\section*{Appendix}

In the body of this paper, we have discussed the leptogenesis scenario
from the universe dominated by $\sneu$. Here, we briefly comment on the
case where the $\sneu$ does not dominate the universe. In this case, the
resultant lepton asymmetry depends on the reheating temperature $T_R$
and the initial amplitude of the oscillation $|\sneu_i|$, and it is
given by the following form~\cite{MY};
\begin{eqnarray}
\label{eq-non-dom}
\frac{n_L}{s} 
&=&
\frac{1}{4}\epsilon_1
\left(
\frac{T_R}{M_1}
\right)
\left(
\frac{|\sneu_i|}{\sqrt{3} M_{pl}}
\right)^2
\nonumber \\
&\simeq&
0.8\times 10^{-10}
\left(
\frac{T_R}{10^7 \GeV}
\right)
\left(
\frac{|\sneu_i|}{M_{pl}}
\right)^2
\left(
\frac{m_{\nu_3}}{0.05 \eV}
\right)
\delta_{\rm eff}
\,.
\end{eqnarray}
Thus, it is possible to produce the desired amount of baryon asymmetry,
avoiding the cosmological gravitino problem, although it depends
crucially on the reheating temperature.

%


%

%
%

\end{document}